\documentclass[preprints,article,accept,moreauthors,pdftex]{Definitions/mdpi} 

\usepackage{stmaryrd}
\usepackage{units}
\usepackage{gensymb}
\usepackage{amssymb}

\usepackage{booktabs} 
\usepackage{multirow}
\usepackage{soul} 
\usepackage{microtype}

\usepackage{upgreek}
\usepackage{textcomp}
\firstpage{1} 
\pubvolume{xx}
\issuenum{1}
\articlenumber{5}
\pubyear{2020}
\copyrightyear{2020}
\history{Received: date; Accepted: date; Published: date}
\pdfoutput=1

\Title{Entropy of a Turbulent Bose-Einstein Condensate}

\Author{Lucas Madeira $^{1,}$*, Arnol Daniel García-Orozco $^{1}$, Francisco Ednilson Alves dos Santos $^2$  and Vanderlei Salvador Bagnato $^{1,3}$}%MDPI: the author names are different from susy, please confirm.
\AuthorNames{Lucas Madeira, Arnol D. Garc\'ia-Orozco and Vanderlei S. Bagnato}

\address{%
$^{1}$ \quad Instituto de F\'isica de S\~ao Carlos, Universidade de S\~ao Paulo, CP 369, S\~ao Carlos, S\~ao Paulo, 13560-970, %Please add post code. (or zip code in the US).
 Brazil; arnolgarcia@ifsc.usp.br (A.D.G.-O.); vander@ifsc.usp.br (V.S.B.)\\
$^{2}$ \quad Departamento de F\'isica, Universidade Federal de S\~ao Carlos, S\~ao Carlos 13565-905, Brazil; santos@ufscar.br\\
$^{3}$ \quad Hagler Fellow, Department of Biomedical Engineering, Texas A\&M University, College~Station,~TX~77843,~USA}

\corres{Correspondence: madeira@ifsc.usp.br}

\abstract{Quantum turbulence deals with the phenomenon of turbulence in quantum fluids, such as superfluid helium and trapped Bose-Einstein condensates (BECs). Although much progress has been made in understanding quantum turbulence, several fundamental questions remain to be answered. In this work, we investigated the entropy of a trapped BEC in several regimes, including equilibrium, small excitations, the onset of turbulence, and a turbulent state. We considered the time evolution when the system is perturbed and let to evolve after the external excitation is turned off.
We derived an expression for the entropy consistent with the accessible experimental data, which is, using the assumption that the momentum distribution is well-known.
We related the excitation amplitude to different stages of the perturbed system, and we found distinct features of the entropy in each of them. In particular, we observed a sudden increase in the entropy following the establishment of a particle cascade. We argue that entropy and related quantities can be used to investigate and characterize quantum turbulence.
}

\keyword{quantum turbulence; Bose-Einstein condensate; out-of-equilibrium; particle cascade}

\begin{document}
%%%%%%%%%%%%%%%%%%%%%%%%%%%%%%%%%%%%%%%%%%
\section{Introduction}

The route to equilibration of a many-body quantum system driven to a far-from-equilibrium state is a question that permeates several areas in physics. Quantum turbulence \cite{Tsatsos2016,Madeira2020} is an example of such out-of-equilibrium systems.
Its classical counterpart, classical turbulence, is a process that occurs in many types of fluids, spanning
the climatic effects that involve large masses down to capillaries.
It is characterized by a large number of degrees of freedom
interacting non-linearly to produce disordered states, both in space and time.
Turbulence presents properties that are universal, regardless of the system under consideration. Many aspects of classical turbulence are not well-understood, so dealing with its quantum version, quantum turbulence, seems very ambitious.
However, turbulence in quantum fluids might be more tractable
than its classical counterpart, because~the vortex circulation
is quantized in the former and continuous for classical fluids.
Additionally, the advances in trapping,~cooling, and tuning the interparticle interactions in atomic Bose--Einstein condensates (BECs) make them excellent candidates for studying quantum turbulence.

Much progress has been made in understanding and characterizing quantum turbulence since the first observation of turbulence in a trapped BEC, and its signature self-similar expansion \cite{Henn2009,Henn2010}.
A~milestone was the identification of an energy cascade
demonstrated by the presence of a power-law in the energy spectrum $E(k)\propto k^{-\alpha}$
\cite{Thompson2013,Navon2016}.

However, there are some intrinsic difficulties in determining the range and exponent of the power-law. The range of length scales and, consequently, momentum scales, available in trapped BEC experiments is very narrow if compared to other systems (superfluid helium, for example).
Additionally, different theoretical models predict exponents there are close together, and experiments do not have yet the necessary precision to distinguish between them. Hence, approaches other than the power-law behavior have been employed to overcome these issues.
Energy and particle fluxes have been used in simulations \cite{Baggaley2014,Marino2020} and experiments \cite{Navon2019,Orozco2020} as alternative methods to investigate and characterize quantum turbulence.

Entropy is a fundamental concept for any physical system. The importance of entropy in determining the macroscopic behavior of a system has been recognized by many scientists, like~Maxwell, Boltzmann, Kelvin, and many others. Conventionally, entropy is determined for a physical system as a function of a parameter, such as temperature, density, or other. Entropy is rarely determined as a general property of the physical system and its properties explored from a broader perspective.
As entropy relates microscopic and macroscopic aspects to systems in equilibrium or tending to equilibrium, it is expected that a system far from its equilibrium state and left to evolve can show, through the determination of its entropy, characteristics of this evolution in search for the re-establishment of equilibrium. Turbulence is one of these cases well-suited for such analysis.
There~are a wide variety of "entropy-like" quantities that can be determined. For a certain physical system, based on the quantities that we can access from measurements, one definition is more~appropriate.

In this work, we investigated the entropy of a trapped BEC in regimes that range from equilibrium up to a turbulent state.
We are interested in the decaying turbulence regime, which is,
the system is perturbed, and we study the time evolution after the external excitations are turned off.
We derived an expression for the entropy that is consistent with our knowledge of the system in order to calculate the entropy using the experimental data that we can acquire. The central assumption was that the values of the momentum distribution are well-known.

Our findings could also be of interest to fields other than quantum turbulence. Several areas would benefit from a better understanding of Bose--Einstein condensates from an entropy point of view, since~isentropic transformations have been widely used to manipulate quantum gases. For~example, the~adiabatic change of the shape of the trap of cold atoms gives the possibility to change the phase-space density in a controlled
manner \cite{Pinkse1997,Kurn1998}.
This led to the proposal to produce BECs using atoms in optical lattices by adiabatically removing the optical lattice \cite{Olshanii2002}.
High-level control of the mode populations of guided-atom lasers was demonstrated by showing that the entropies per particle of an optically guided-atom laser and the one of the trapped BEC from which it has been produced are the same
\cite{Gattobigio2009}.
In the context of two-component fermionic gases, the adiabatic tuning of the interspecies scattering length from positive to negative
values lead to the reversible formation of a molecular BEC
\cite{Carr2004,Williams2004,Bourdel2004,Partridge2005,Zwierlein2005}.

This work is organized, as follows. In Section~\ref{sec:experimental}, we summarize the experimental procedure of producing a BEC, the introduction of an external excitation, and measuring the momentum distribution.
The theoretical aspects of this work are presented in Section~\ref{sec:theory}. We derive an expression for the entropy consistent with our knowledge of the system, and we show that it agrees with well-known results and limits.
Section~\ref{sec:results} contains our results.
In Section~\ref{sec:momentum}, we show how to reconstruct the three-dimensional momentum distribution of the cloud from the experimental data, and we relate the excitation amplitude to different stages of the system: small perturbations, the onset of turbulence, and the turbulent regime.
The momentum distributions are used in Section~\ref{sec:entropy} to compute the entropy for these different regimes. Finally, we present our conclusions and outlook in Section~\ref{sec:conclusions}.

%%%%%%%%%%%%%%%%%%%%%%%%%%%%%%%%%%%%%%%%%%
\section{Experimental Procedure}
\label{sec:experimental}

The initial phase of the experiment consists in the production of a Bose-Einstein condensate, containing approximately $4\times 10^5$ $^{87}$Rb atoms in the hyperfine state
$|F, m_F \rangle=| 2,2 \rangle$, confined in a Quadrupole--Ioffe configuration (QUIC) magnetic trap of frequencies
$\omega_{r}/ 2\pi = \unit[237.3(8)]{Hz}$ and $\omega_{x} / 2\pi = \unit[18.7(2)]{Hz}$.
The unperturbed BEC has a condensate fraction of $70(5)\%$, chemical potential $\upmu_0/k_{B} = \unit[124(5)]{nK}$, and healing length $\xi_0 = 0.15(2)$ $\upmu$m.
The number density is of $n=10^{14}$ cm$^{-3}$, which results in $na^3=10^{-5}$, $a$ being the $s$-wave scattering length of $^{87}$Rb.
The critical temperature for this trapping geometry is $T_c=$ 430 nK.
The temperature of the unperturbed sample is approximately constant for the time intervals reported in this work, 220(15) nK, corresponding to $0.5 T_c$.
Details of the experimental procedure and other technical remarks can be found in previous works \cite{Henn2010,Seman2011,Shiozaki2011}.

After the condensate is produced, while it is still in the trap, an oscillating magnetic field is applied, thus driving the BEC out-of-equilibrium. The excitation potential corresponds to
\begin{equation}
V_{\rm exc}(\mathbf{r},t)= A\big[1-\cos{(\Omega t)}\big]x/R_x,
\end{equation}
where $R_x$ = 42 $\upmu$m is the in-trap extent of the BEC along the $x$-axis of the trap.
The field is produced by a pair of anti-Helmholtz coils placed with their axis tilted by a small angle, of $\approx$5$^{\circ}$
, with respect to the axis of the trap.
Because the coils are not aligned with the condensate axis, the oscillations generate deformations, displacements, and rotations in the cloud. The amplitude, time, and frequency of the disturbances can be varied.
In this work, we keep the excitation frequency fixed at $\Omega/(2\pi)=$ 230.5 Hz, close to the radial trapping frequency.
The excitation amplitude is varied from small values until an amplitude is reached where the momentum distribution corresponds to a non-equilibrium state with turbulent characteristics.
The excitation time may also vary, because there is a compromise between the excitation application time and the amplitude to generate the turbulent state. For example, larger amplitudes need less time to reach similar conditions. The range of amplitudes to obtain turbulence was the topic of investigation in previous works \cite{Henn2010,Seman2011,Shiozaki2011}.

The excitation is applied during a time $t_{\rm exc} = 5\tau$, where $\tau = 2\pi/\Omega$. The amplitude $A$ is varied, ranging from from $0$ (no perturbation) to $0.60 \upmu_0$.~This protocol drives a BEC in initial thermal equilibrium to an out-of-equilibrium state. Subsequently, the atomic cloud is held for a time $t_{\rm hold}$ inside the trap, which we vary from 20 to 90 ms, leading to the temporal evolution of the momentum~distribution.

As the amplitude of the external excitation increases, so does the temperature of the cloud, the~atom losses, and the depletion of the condensate.
At $t_{\rm hold}=20$ ms, the number of atoms is close to 
$4\times 10^5$ for all amplitudes.
For the longest holding times that we investigated, $t_{\rm hold}=$ 90 ms, the~atom loss never exceeds 20\% for $0.20 \upmu_0\leqslant A \leqslant 0.50\upmu_0$, and it is much more pronounced for the highest amplitude considered, 50\%.
The temperature of the cloud is $\sim$0.5 $T_c$ at $t_{\rm hold}=20$ ms, while, at~$t_{\rm hold}=90$ ms it is close to $0.7 T_c$ for $0.20 \upmu_0\leqslant A \leqslant 0.40\upmu_0$ and approximately $T_c$ for $A \geqslant 0.50\upmu_0$.
Finally, the condensate fraction is approximately 40\% at $t_{\rm hold}=20$ ms (if the excitation protocol is employed) and it can drop to as low as 20\% for the highest amplitude and $t_{\rm hold}=90$ ms.

To probe the state of the gas after a $t_{\rm hold}$, we turn off the trap potential and measure the momentum distribution $n(k, t)$ using absorption images taken from the ballistic expansion of the cloud after a time of flight (TOF) of $t_{\rm TOF}=\unit[30]{ms}$.
Following the release of the trap, the distance that an atom has traveled from its center is given by $r=\hbar t_{\rm TOF}k/m$, where $\hbar$ is Planck's constant and $m$ is the atomic mass.
Hence, the TOF procedure corresponds to a Fourier transform of the spatial distribution into the momentum distribution,
\begin{equation}
n(r)\propto n\left(\frac{\hbar t_{\rm TOF}k}{m}\right).
\end{equation}

The expansion of a dilute atomic cloud is relatively well-understood in the regime of momenta smaller than the
inverse of the healing length \cite{Pitaevskii2016}.~In this regime, schemes based on scaling transformations \cite{Kagan1996,Castin1996}
or on the hydrodynamic picture \cite{Dalfovo1997}
have proven to be very successful in describing the dynamics of the expansion.

However, the description of the expansion at a microscopic
level involves knowledge of high-momenta components of the momentum distribution, which is more challenging.
For example, in Ref.~\cite{Qu2016} the authors
studied the expansion of a weakly interacting Bose gas at zero temperature, following its release from an isotropic three-dimensional harmonic trap. They calculated the time dependence of the momentum
distribution with a special focus on the behavior of the contact parameter, which depends on the tail of the distribution.
They showed that the momentum distribution changes dramatically if the interaction is present during the expansion, and that the high momentum tail decreases and eventually disappears for large $t_{\rm TOF}$.

In the presence of interactions, quantum fluctuations
deplete the condensate, thus inducing correlations in TOF measurements.
Numerical methods have been used to compute momentum correlations in a two-dimensional harmonically trapped interacting Bose gas at zero
temperature during the TOF procedure \cite{Lovas2017}.
It was found that the momentum distribution is dominated by a central peak, due to the condensate, and the amplitude of the normal fraction of the gas is overwhelmed by this central peak. However, the latter contribution is much more extended in the momentum space, represented by a slowly decaying tail.

These shortcomings of the TOF technique do not significantly impact our results. A common assumption of the investigations concerning the limitations of the TOF measurements is that the interaction energy is dominant in these systems.
However, the turbulent state is kinetically dominated,
which means that the interaction energy plays a small role in the effects
of a turbulent cloud \cite{Caracanhas2013}.
Hence, this technique has been used successfully to obtain the momentum distribution of turbulent trapped BECs \cite{Thompson2013,Navon2016}.

%%%%%%%%%%%%%%%%%%%%%%%%%%%%%%%%%%%%%%%%%%
\section{Extracting the Entropy from Experimental Data}
\label{sec:theory} 

The momentum distribution $n(k)$ corresponds to information about the
square of the amplitude of the condensate's wave function, while its phase cannot be determined in these experiments.
However, we can derive an expression for the entropy using the assumption that the momentum distribution values are well-known, which is what we can obtain from the experimental data.
In the following sections, we derive the formula that we use to calculate the entropy of the turbulent BECs. We show that the expression is consistent with well-known statistical physics results. 

\subsection{Classical Treatment}
\label{sec:classical}

First, let us start with some remarks about the notation. We work in momentum space, with~coordinates given by $\mathbf{k}_{j}$,
where the index $j$ identifies the different points in momentum space,
and~we employ the shorthand notation $\psi(\mathbf{k}_{j})\equiv\psi_{j}$ for the fields.
The functional ``volume'' element used in the integrations is given by
\begin{equation}
D\{\psi^{\ast},\psi\}\equiv\prod_{\mathbf{k}_{j}}d\psi^{\ast}(\mathbf{k}_{j})d\psi(\mathbf{k}_{j})\equiv\prod_{j}d\psi_{j}^{\ast}d\psi_{j}.
\end{equation}

The fields can be separated into real and imaginary parts, $\psi_j=R_j+iI_j$,
with a differential given by $d\psi_j^{\ast}d\psi_j\equiv dR_jdI_j$ \cite{Kleinert2009}.
We use the Mandelung transformation to write the fields as $\psi_j=A_j e^{i\varphi_j}$,
with $R_j=A_j\cos\varphi_j$ and $I_j =A_j\sin\varphi_j$.
Hence, the integrations can be done in ``polar'' coordinates,
$d\psi_j^{\ast}d\psi_j\equiv dA_j\ d\varphi_j\ A_j$.

The entropy can then be defined according to:
\begin{equation}
\label{eq:S}
S=-\int D\{\psi^{\ast},\psi\}\rho\{\psi^{\ast},\psi\}\ln\left(\rho\{\psi^{\ast},\psi\}\right).
\end{equation}

It is important to notice that additive and multiplicative constants
could be added to this definition. The Principle of Maximum Entropy states that the probability distribution
that best represents the knowledge of the system is the one that maximizes
Equation~(\ref{eq:S}) obeying the required constraints. Such~constraints are typically included via Lagrange multipliers.

As stated before, the experimental data corresponds to well-defined values of the density ${n}(\mathbf{k}_j)\equiv n_j=\langle \psi^{\ast}_j \psi_j \rangle$, which corresponds to the first constraint, while a second constraint must be introduced to ensure that the integral of the probability density function is equal to one.
Hence, we want to maximize the entropy with these constraints enforced
by the Lagrange multipliers $\upmu_j$ and $\lambda$, 
\begin{equation}
\frac{\delta}{\delta\rho\{\psi^{\prime\ast},\psi^{\prime}\}}\left[S+\lambda\int
D\{\psi^{\ast},\psi\}\rho\{\psi^{\ast},\psi\}
+\sum_{j}\upmu_{j}\int D\{\psi^{\ast},\psi\} \rho\{\psi^{\ast},\psi\}\psi_{j}^{\ast}\psi_{j}\right]=0.
\end{equation}

The equation above leads to the probability density function
\begin{equation}
\label{eq:rho}
\rho\{\psi^{\ast},\psi\}=\prod_{j}\frac{\upmu_{j}}{\pi}e^{-\upmu_{j}\psi_{j}^{\ast}\psi_{j}}.
\end{equation}

The expectation values of the density can be obtained by performing the Gaussian integrations,
\begin{equation}
\label{eq:muj}
n_j=\left\langle \psi_{j}^{\ast}\psi_{j}\right\rangle =\int D\{\psi^{\ast},\psi\}\prod_{k}\frac{\upmu_{k}}{\pi}e^{-\upmu_{k}\psi_{k}^{\ast}\psi_{k}}\psi_{j}^{\ast}\psi_{j}=\frac{1}{\upmu_{j}}.
\end{equation}

Substituting Equations~(\ref{eq:rho}) and (\ref{eq:muj}) into Equation~(\ref{eq:S}) yields
\begin{eqnarray}
S&=&-\int D\{\psi^{\ast},\psi\}
\prod_{j}\frac{1}{\pi n_j}e^{-\psi_{j}^{\ast}\psi_{j}/n_j}
\ln\left(
\prod_{k}\frac{1}{\pi n_k}e^{-\psi_{k}^{\ast}\psi_{k}/n_k}
\right)\nonumber\\
&=&-\ln\left[\prod_{j}\frac{1}{\pi n_j}\right]+\sum_{\mathbf{k}}1=C+\sum_{j}\ln\left(n_{j}\right).
\end{eqnarray}

The additive constant $C$ can be neglected, so that we can define
\begin{equation}
\label{eq:Sc}
S_c=\sum_{j}\ln\left(n_{j}\right).
\end{equation}

Notice that, since usually $n(\mathbf{k})\rightarrow 0$ as $k\rightarrow\infty$,
Equation~(\ref{eq:Sc}) may give rise to ultraviolet divergences, which
is typical of classical treatments.
In the following section, we show how this issue goes away with quantum treatment.

Finally, it is possible to show that the entropy of Equation~(\ref{eq:Sc}) agrees with a thermal distribution for $n(\mathbf{k})$. If we impose the constraint on the number of particles and energy ($\sum_j n_j \hbar^2k_j^2/(2m)$) through the Lagrange
multipliers $\lambda_1$ and $\lambda_2$, the extremization of the momentum distribution yields
\begin{equation}
\frac{\partial}{\partial n_{j}}\left[\sum_{l}\ln\left(n_{l}\right)-\lambda_{1}\sum_{l}n_{l}-\lambda_{2}\sum_{l}\frac{\hbar^2 k_l^{2}}{2m} n_{l}\right]=0.
\end{equation}

This leads to
\begin{equation}
n(\mathbf{k})=\frac{1/\lambda_{2}}{\frac{\hbar^2k^{2}}{2m}+\lambda_{1}/\lambda_{2}},
\end{equation}
which is the thermal Rayleigh-Jeans distribution with
$k_B T=1/\lambda_2$ and $\upmu=\lambda_1/\lambda_2$, where $T$ is the temperature and $\upmu$ is the chemical potential. 

\subsection{Quantum Treatment}
\label{sec:quantum}

For quantum systems, we must consider the von Neumann entropy,
\begin{equation}
\label{eq:vonN}
S=-\text{Tr}\left[\hat{\rho}\ln\ \hat{\rho}\right],
\end{equation}
where $\hat{\rho}$ is the density matrix.
In order to relate this formalism with the experimental data, we~consider
the constraints $n_j\equiv {n}(\mathbf{k}_j)=\left\langle \hat{\psi}^{\dagger}(\mathbf{k}_j)\hat{\psi}(\mathbf{k}_j)\right\rangle\equiv \left\langle \hat{\psi}_j^{\dagger}\hat{\psi}_j\right\rangle$. Differently from the previous section, the~operators obey the canonical commutation relations, $\left[\hat{\psi}_{j},\hat{\psi}_{k}^{\dagger}\right]=\delta_{jk}$.

When considering these assumptions, the density matrix is given by
\begin{eqnarray}
\label{eq:dmatrix}
\hat{\rho} &=& Z^{-1}e^{-\sum_{j}f_{j}\hat{\psi}_{j}^{\dagger}\hat{\psi}_{j}},
\nonumber\\
Z &=& \prod\limits_{j}\frac{1}{1+e^{-f_{j}}},
\end{eqnarray}
with
\begin{equation}
\label{eq:nj_fj}
n_{j}=\frac{e^{-f_{j}}}{1-e^{-f_{j}}}.
\end{equation}

Now, the entropy becomes
\begin{align}
\label{eq:S_fj}
S =-\text{Tr}\left[Z^{-1}e^{-\sum_{j}f_{j}\hat{\psi}_{j}^{\dagger}\hat{\psi}_{j}}\ln\left(Z^{-1}e^{-\sum_{j}f_{j}\hat{\psi}_{j}^{\dagger}\hat{\psi}_{j}}\right)\right]=-\sum_{j}\ln\left(1-e^{-f_{j}}\right)+\sum_{j}\frac{f_{j}n_{j}}{1-e^{-f_{j}}}.
\end{align}

Combining Equations~(\ref{eq:nj_fj}) and (\ref{eq:S_fj}) yields an expression for the entropy in terms of only the momentum distribution $n_j$,
\begin{equation}
\label{eq:entropy}
S=\sum_{j}\left[\left(1+n_{j}\right)\ln\left(1+n_{j}\right)-n_{j}\ln\ n_{j}\right].
\end{equation}

Although we discussed thermal equilibrium, Equation~(\ref{eq:entropy}) was derived without assuming it. The only assumption was that the expected values of the momentum distribution are well-known, which is valid both for systems in and out of equilibrium.
This formula does not suffer from the same ultraviolet issues that the one found with the classical treatment, Equation~(\ref{eq:Sc}), since
\begin{equation}
\left(1+n_{j}\right)\ln\left(1+n_{j}\right)-n_{j}\ln\ n_{j}\overset{n_{j}\shortrightarrow 0}{\longrightarrow}0.
\end{equation}

Additionally, for large occupation numbers we recover the classical formula,
\begin{equation}
\left(1+n_{j}\right)\ln\left(1+n_{j}\right)-n_{j}\ln\ n_{j}\overset{n_{j}\shortrightarrow\infty}{\longrightarrow}\ln\ n_{j}.
\end{equation}

Finally, as a consistency check,
we can apply Equation~(\ref{eq:entropy}) to non-interacting bosons in equilibrium to recover the Bose--Einstein distribution.
If we impose the constraints on the number of particles and the energy,
via Lagrange multipliers $\lambda_1$ and $\lambda_2$, 
\begin{equation}
\frac{\partial}{\partial n_{j}}\left[\sum_{l}\left[\left(1+n_{l}\right)\ln\left(1+n_{l}\right)-n_{l}\ln\ n_{l}\right]-\lambda_{1}\sum_{l}n_{l}-\lambda_{2}\sum_{l} \frac{\hbar^2 k_l^{2}}{2m} n_{l}\right]=0.
\end{equation}

After a few manipulations, we get
\begin{equation}
n(\mathbf{k})=\frac{1}{e^{\lambda_{1}+\lambda_{2} \frac{\hbar^2 k^{2}}{2m}}-1},
\end{equation}
which is the thermal Bose--Einstein distribution with $\lambda_{1}=-\upmu/(k_B T)$ and $\lambda_{2}=1/(k_B T)$.
We should note that this distribution should not be used to describe the data presented in the following section, since it corresponds to an interacting system driven out-of-equilibrium (for $A\neq 0$).

%%%%%%%%%%%%%%%%%%%%%%%%%%%%%%%%%%%%%%%%%%
\section{Results}
\label{sec:results}

\subsection{Momentum Distribution}
\label{sec:momentum}

\begin{figure}[H]
\centering
\includegraphics[width=0.7\linewidth]{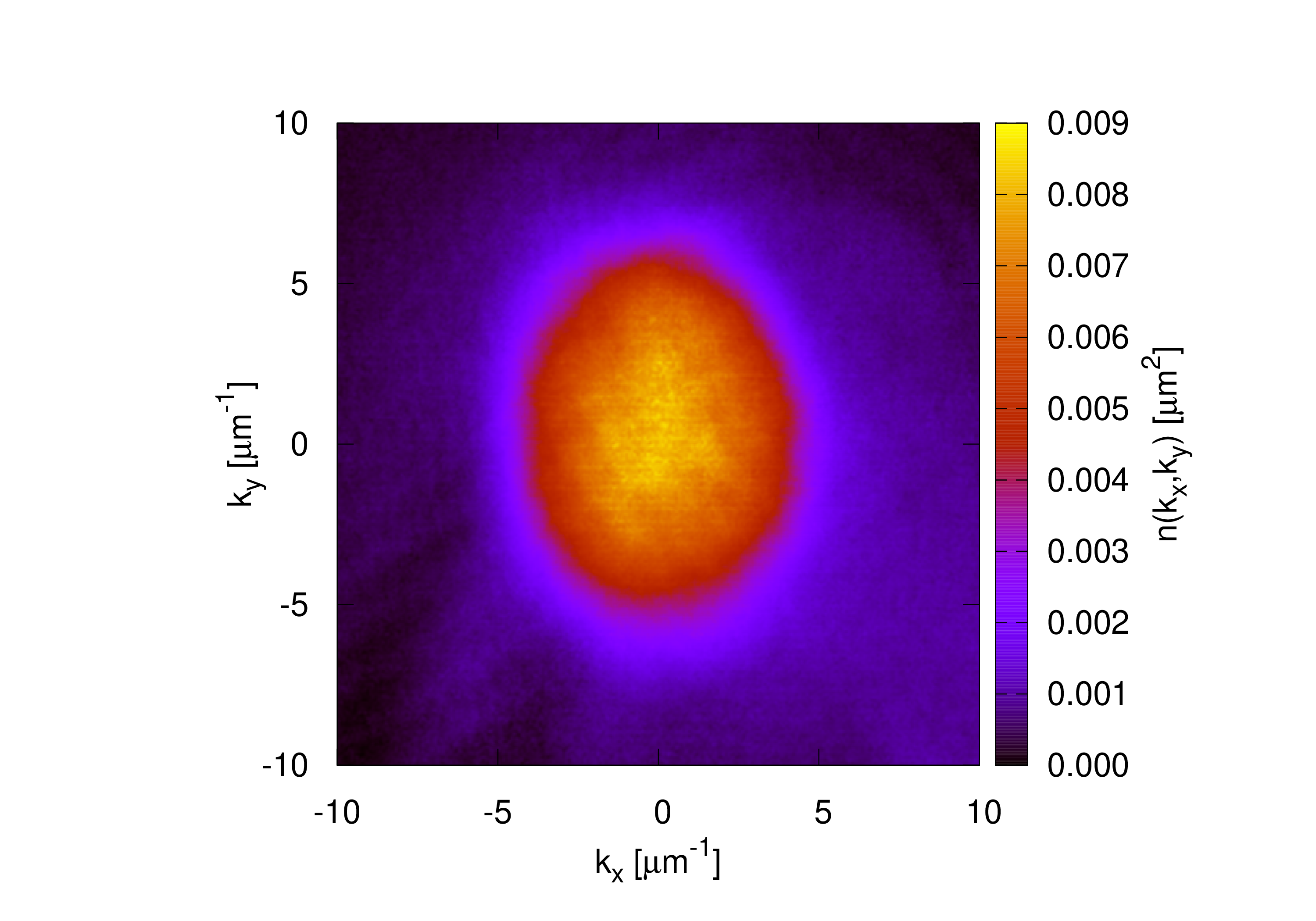}
\caption{
Momentum distribution of an unperturbed cloud ($A=0$) held in the trap for a short time ($t_{\rm hold}=$ 20 ms) obtained through an optical absorption image following the release of the trap.
The normalization is such that the integration over the plane yields one.
}
\label{fig:cloud}
\end{figure}

\begin{figure}[H]
\centering
\includegraphics[angle=-90,width=0.45\linewidth]{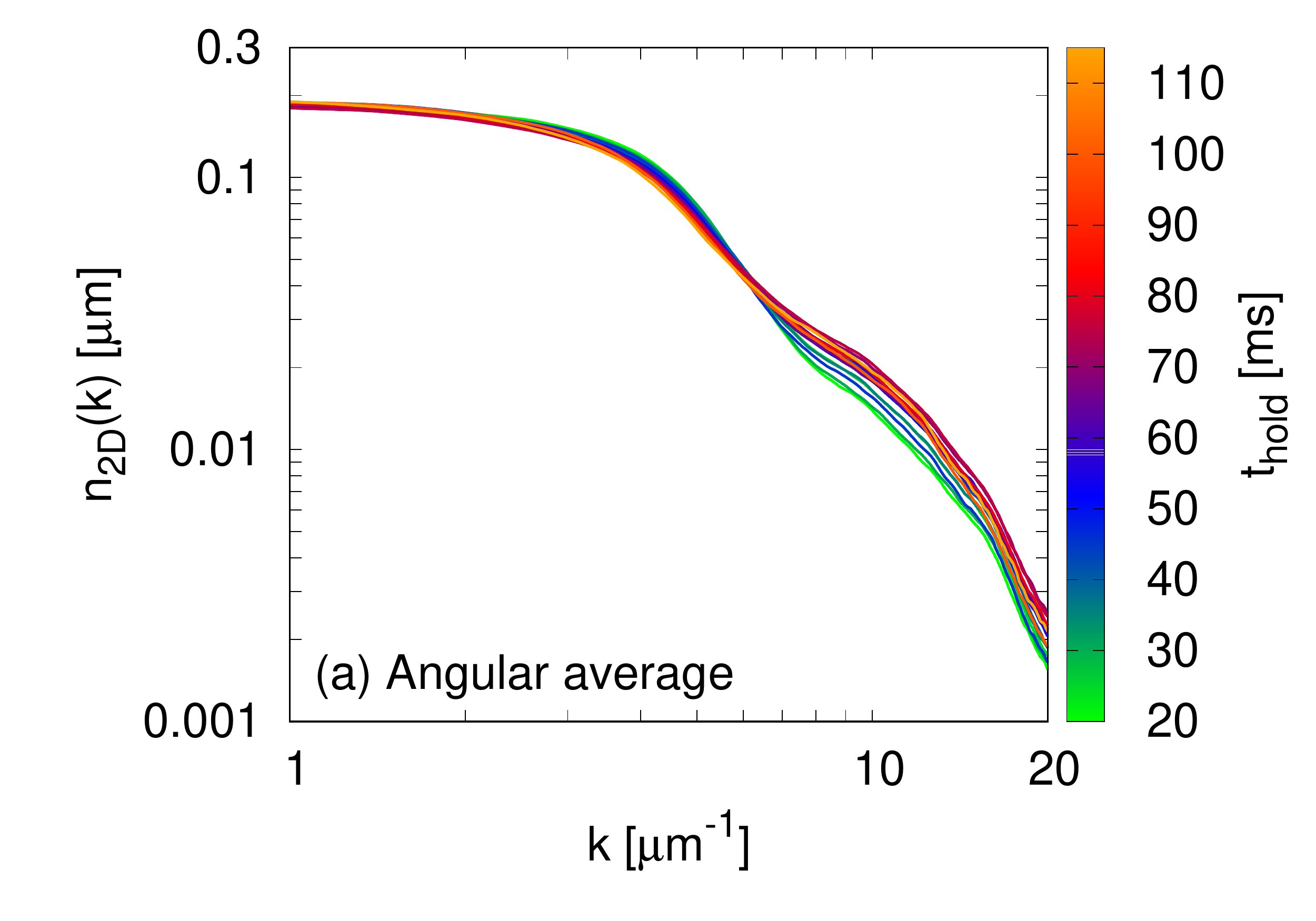}
\includegraphics[angle=-90,width=0.45\linewidth]{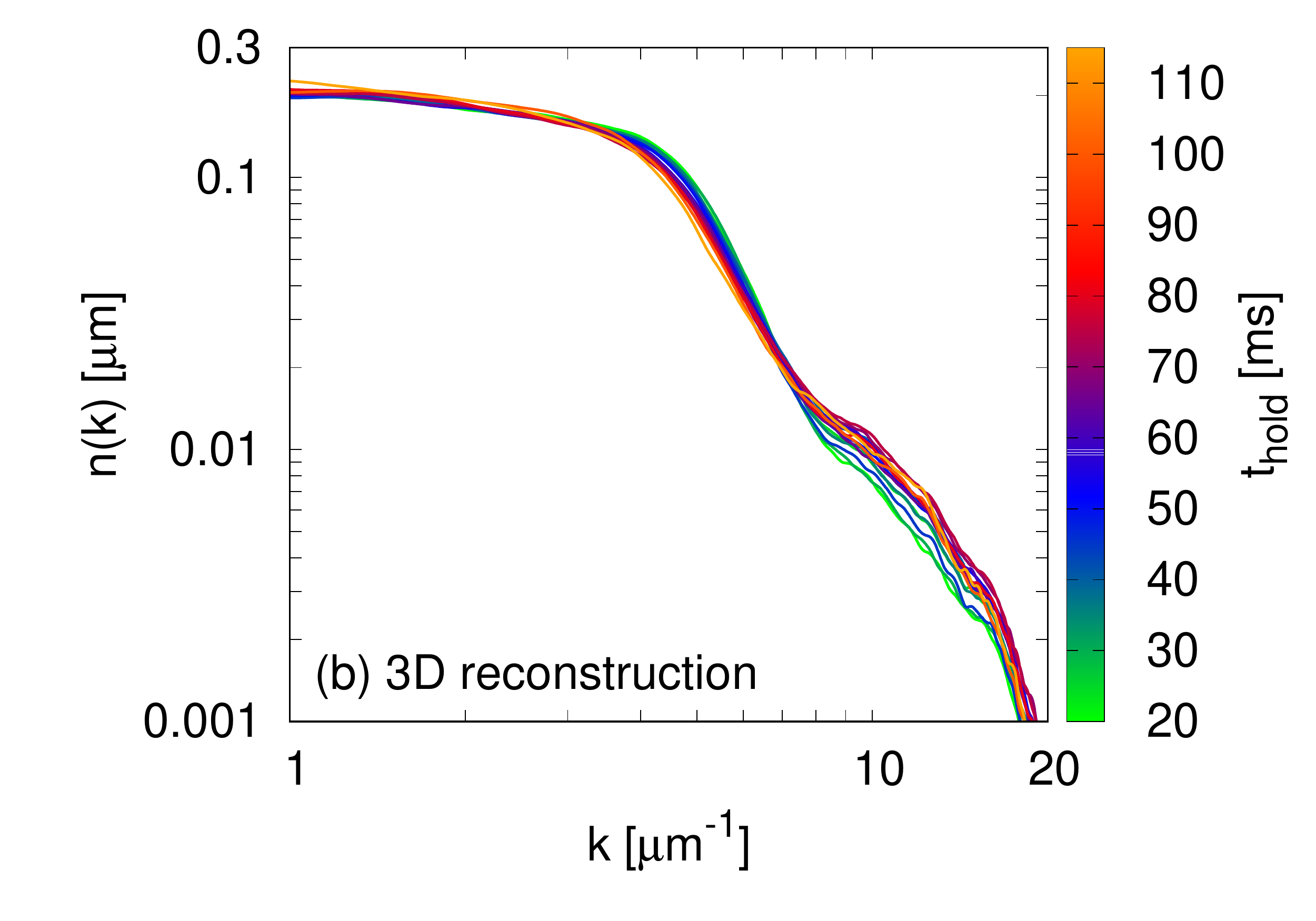}
\caption{Momentum distribution
of an unperturbed cloud ($A=0$) held in the trap for different times ($t_{\rm hold}=$) from 20 to 115 ms. We show both (\textbf{a}) the angular average using the absorption images, and (\textbf{b}) the three-dimensional reconstruction of the momentum distributions using the inverse Abel transform, Equation~(\ref{eq:abel}).
The profiles for different hold times are very similar, with small differences at high-momenta due to heating in the system. 
}
\label{fig:nk_0}
\end{figure}

The optical absorption image of the cloud after the TOF produces an image on the plane $(k_x, k_y)$, as illustrated in Figure~\ref{fig:cloud}.
The momentum distribution is obtained by averaging the values within the
interval $k=\sqrt{k_x^2+k_y^2}$ and $k+\delta k$, with $\delta k \approx 0.04$ $\upmu$m$^{-1}$.
In the left panel of Figure~\ref{fig:nk_0}, we~show the result of this procedure for $A=0$. However, the absorption image corresponds to a two-dimensional projection of the cloud, and we would like to work with the three-dimensional distribution. A~procedure that has been successfully used in the past to overcome this difficulty~\cite{Thompson2013,Navon2016} consists in employing the inverse Abel transform \cite{Hickstein2019}, which reconstructs the three-dimensional distribution $n(k)$ from its two-dimensional projection $n_{\rm 2D}(k)$ according to
\begin{equation}
\label{eq:abel}
n(k)=-\frac{1}{\pi}\int_{k}^\infty \frac{d n_{\rm 2D}(k')}{dk'}\frac{dk'}{\sqrt{k'^2-k^2}}.
\end{equation}

In the right panel of Figure~\ref{fig:nk_0}, we show the result of applying this transformation to the two-dimensional momentum distributions. For the remainder of this paper, we will only employ the reconstructed three-dimensional momentum distributions.
\vspace{-12pt}

\begin{figure}[H]
\centering
\includegraphics[angle=-90,width=0.4\linewidth]{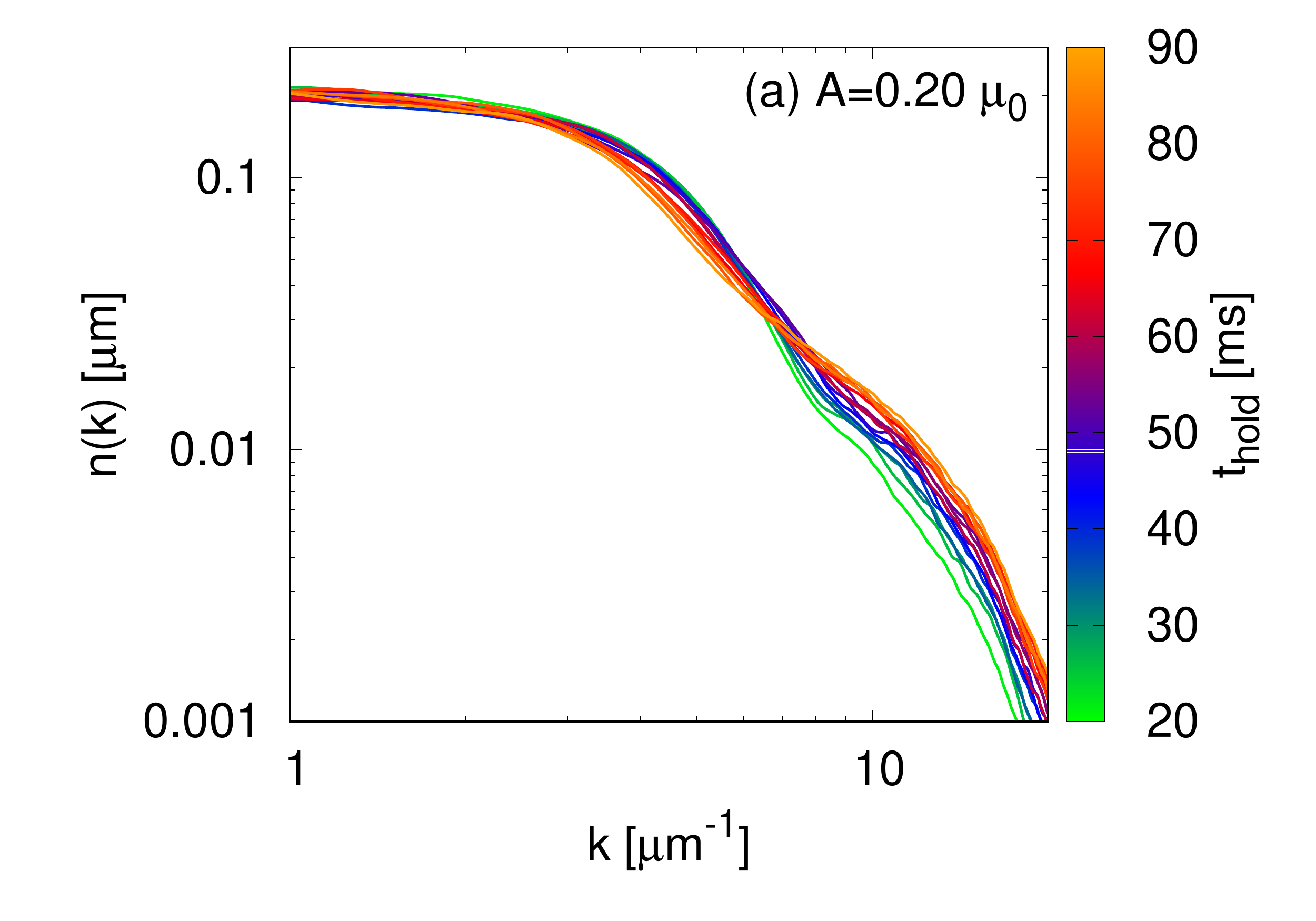}
\includegraphics[angle=-90,width=0.4\linewidth]{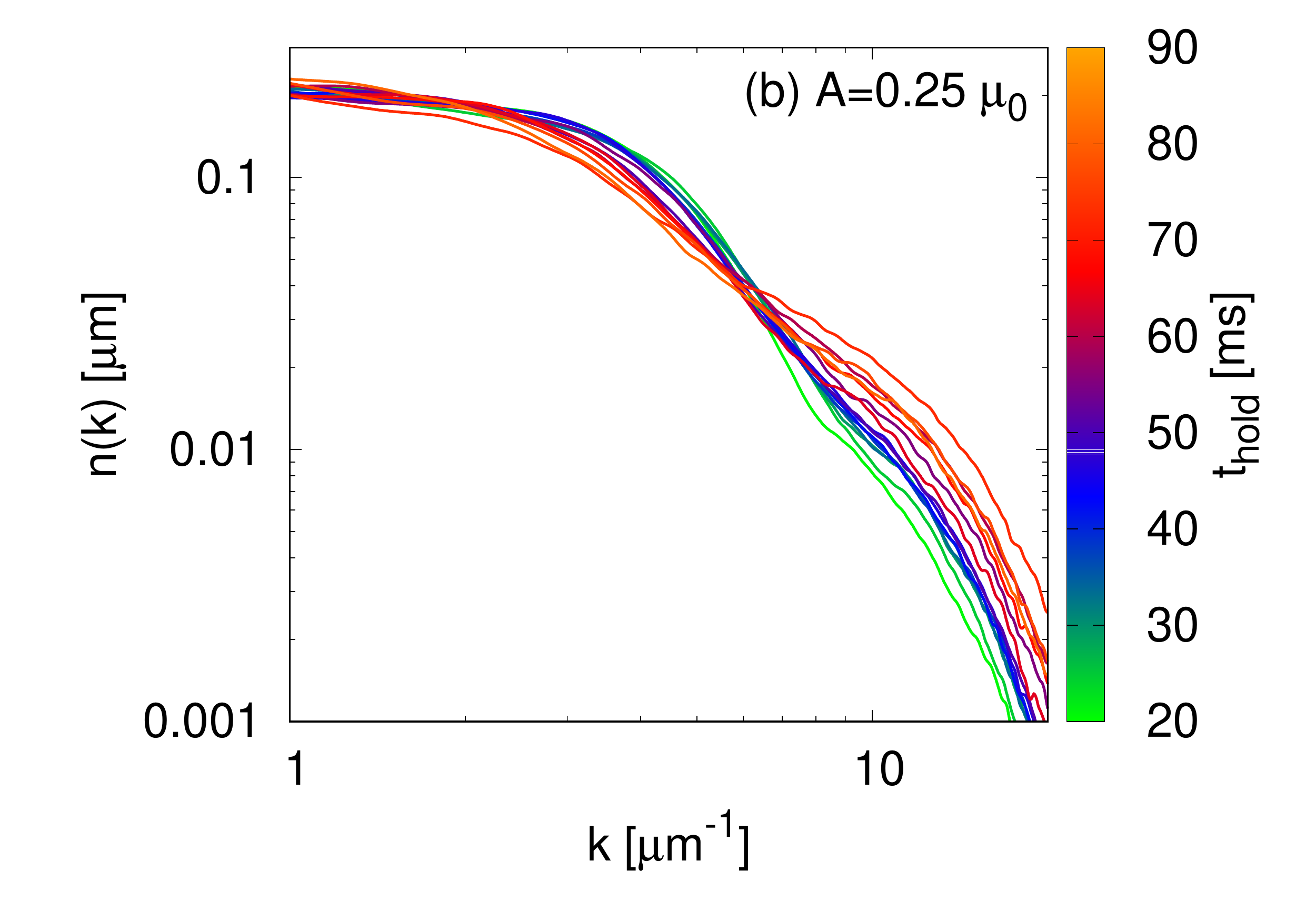}
\includegraphics[angle=-90,width=0.4\linewidth]{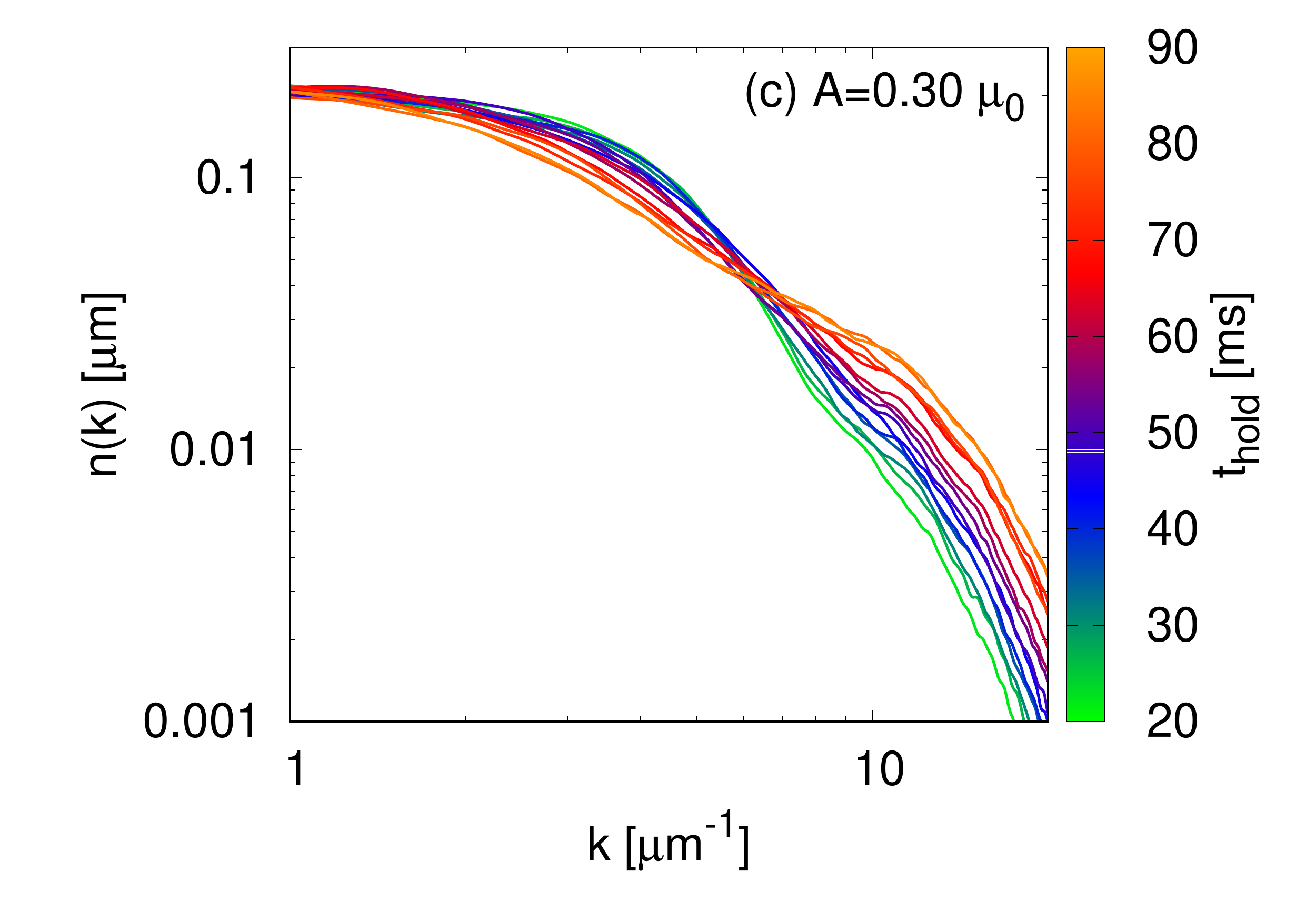}
\includegraphics[angle=-90,width=0.4\linewidth]{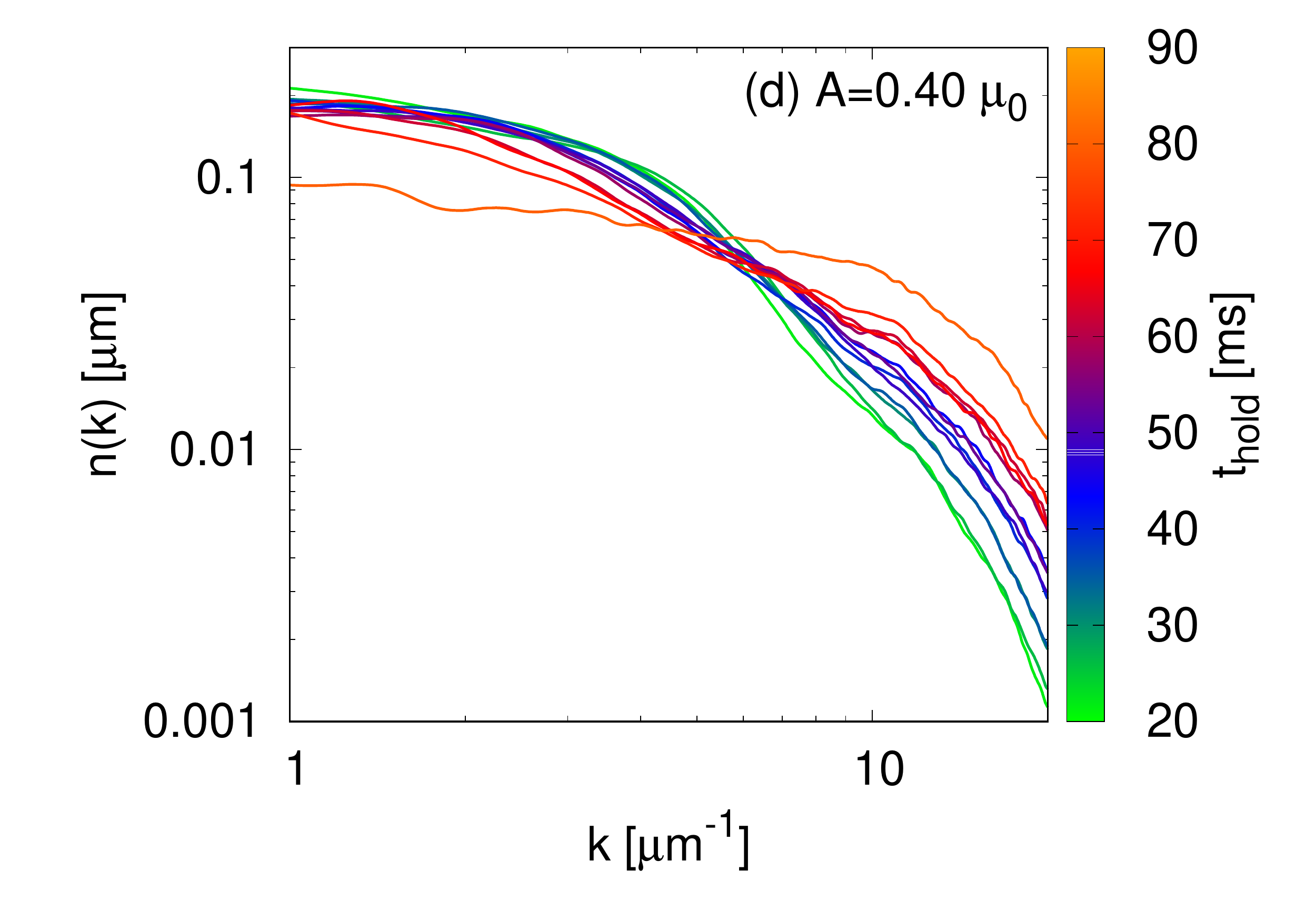}
\includegraphics[angle=-90,width=0.4\linewidth]{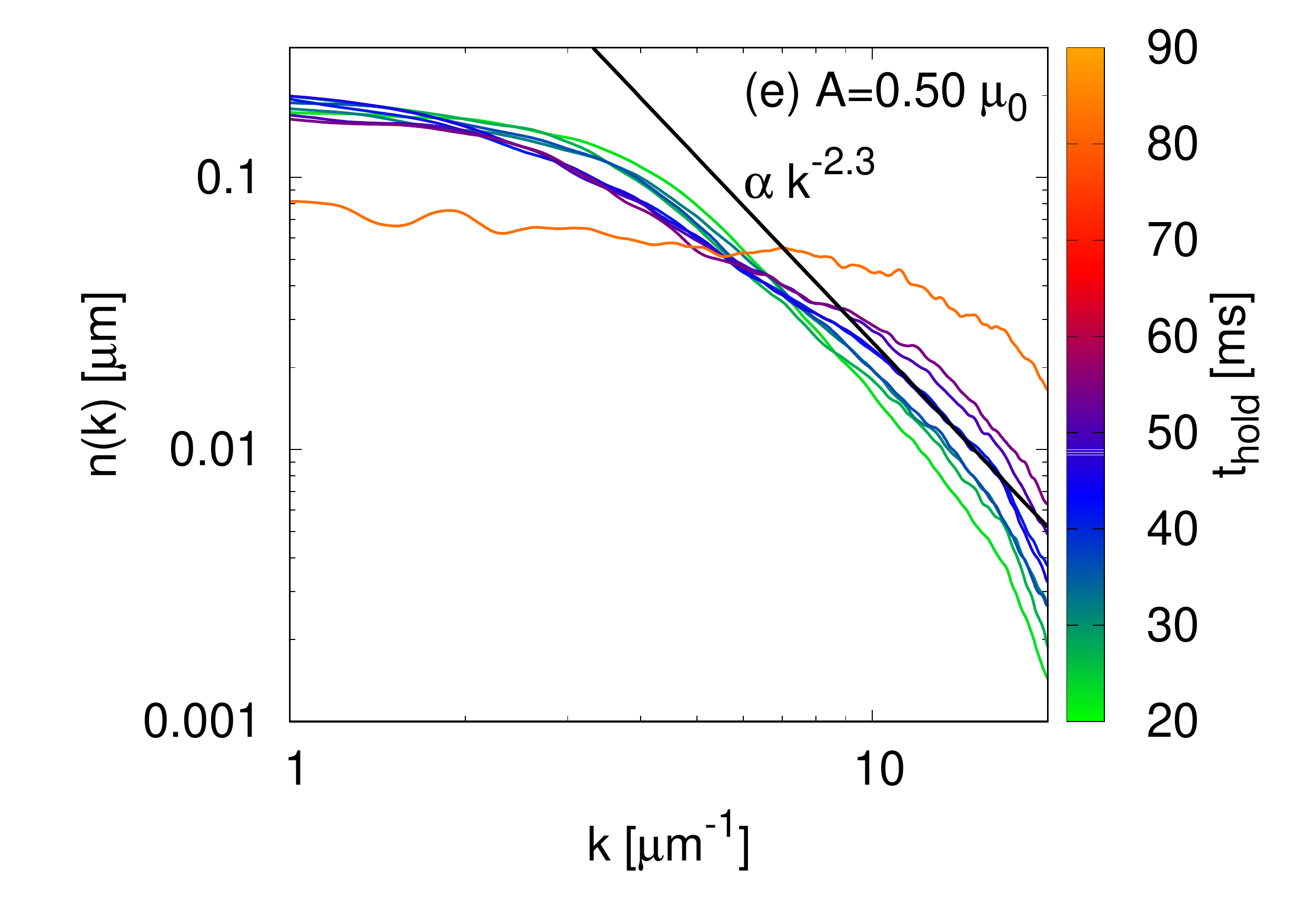}
\includegraphics[angle=-90,width=0.4\linewidth]{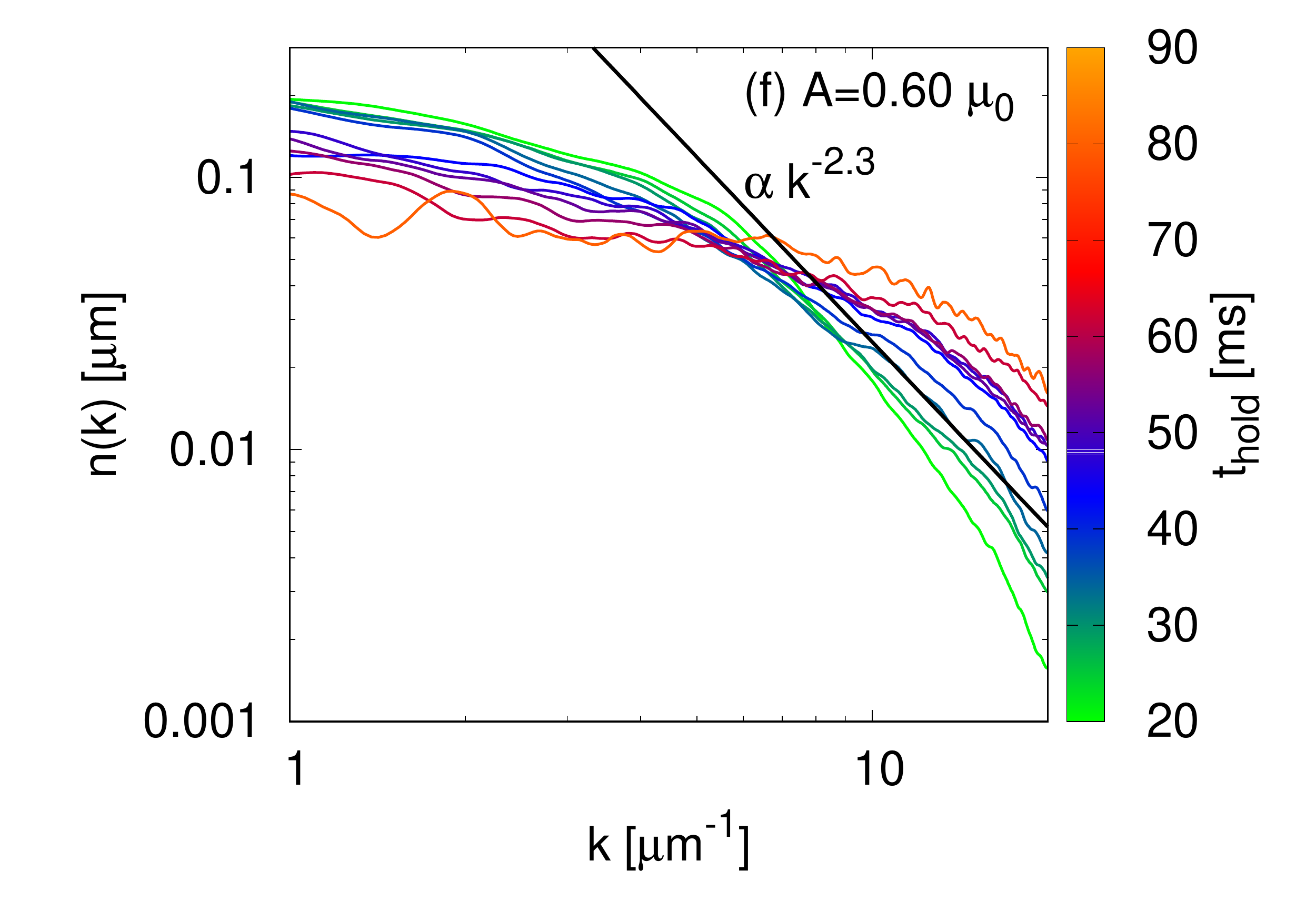}
\caption{Momentum distribution of the cloud for the excitation amplitudes
$A=0.20$, 0.25, 0.30, 0.40, 0.50, and 0.60$\upmu_0$, panels (\textbf{a}--\textbf{f}) respectively,
and holding times ranging from 20 to 90 ms.
Increasing the excitation amplitude corresponds to larger energy input to the BEC, thus driving the system toward higher-momenta regions.
For strong enough excitations, $A\geqslant$ 0.50$\upmu_0$ in this experimental setting, the system enters a turbulent regime with a particle cascade characterized by a power-law $n(k) \propto k^{-\delta}$. In panels (\textbf{e},textbf{f}), we plot a line corresponding to $\propto k^{-2.3}$ to guide the eye.
}
\label{fig:nk}
\end{figure}

In Figure~\ref{fig:nk}, we show the momentum distributions for non-zero perturbations for different holding times. For the lowest excitation amplitude, $A=0.20$ $\upmu_0$, the profile is
similar to the one of the unperturbed cloud, Figure~\ref{fig:nk_0}b.
The momentum distributions of $A=0.25$ and 0.30$\upmu_0$
show the system migrating to higher-momenta, as the energy input of the excitation is increased.
Figure~\ref{fig:nk}d depicts a slightly higher excitation, $A=0.40\upmu_0$, placing the system in a regime that we would best describe as the onset of turbulence.
For strong enough amplitudes, in this particular experimental setting $A\geqslant$ 0.50$\upmu_0$, the system is in a turbulent state with a particle cascade characterized by a power-law behavior $n(k) \propto k^{-\delta}$.
In the region 10$\upmu$m$^{-1}$ $\leqslant k \leqslant 17$ $\upmu$m$^{-1}$
we observe such behavior with $\delta=2.3(2)$ for $t_{\rm hold}$ close to 35 ms. In Figure~\ref{fig:nk}e,f, we plot this power-law to guide the eye. Finally, for long hold times, the amplitudes $A=0.40$, 0.50, and 0.60$\upmu_0$ display momentum distributions compatible with the thermalization of the cloud.

\vspace{-6pt}

\subsection{Entropy}
\label{sec:entropy}

Finally we can compute the entropy using Equation~(\ref{eq:entropy}).
We should note that the profiles of Section~\ref{sec:momentum} were normalized, such that their integral yields one, to make comparisons easier.
However, the derivation of Equation~(\ref{eq:entropy}) assumes the normalization $\sum_i n_i=N$, the total number of atoms. So, we employ the latter normalization and report the entropy per particle for the following entropy~calculations.

\begin{figure}[H]
\centering
\includegraphics[angle=-90,width=0.7\linewidth]{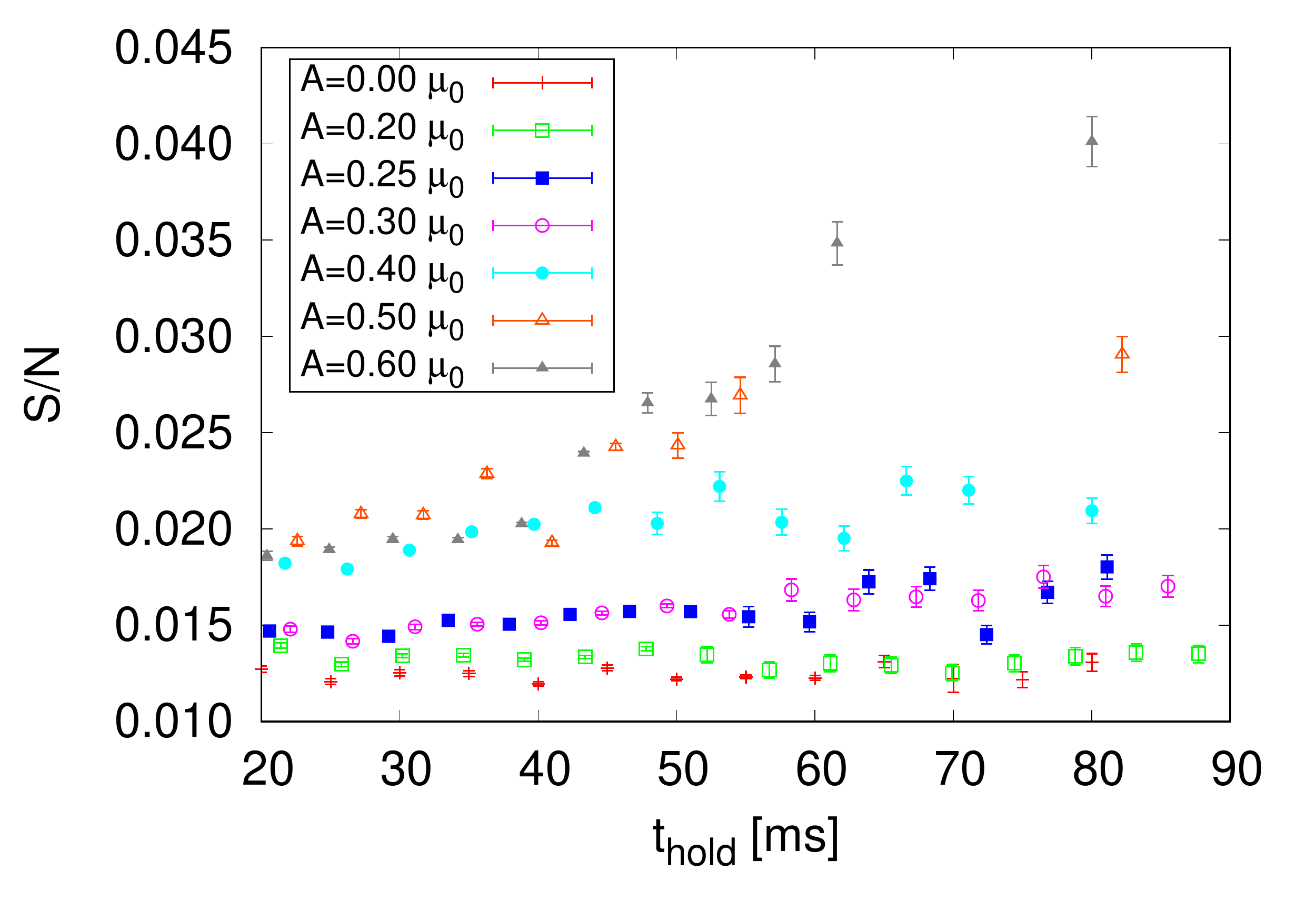}
\caption{
Entropy per particle calculated using Equation~(\ref{eq:entropy}) for several excitation amplitudes as a function of the time held in the trap.
The unperturbed BEC corresponds to a constant entropy per particle over time. Increasing the amplitude corresponds to higher values of the entropy per particle until the turbulent regime is reached, $A\geqslant$ 0.50$\upmu_0$. The particle cascade that occurs at $\approx$ 35 ms (see~Section~\ref{sec:momentum}) is accompanied by a sudden increase in the entropy per particle, as seen for $t_{\rm hold}\geqslant$~45~ms.
}
\label{fig:entropy}
\end{figure}

Figure~\ref{fig:entropy} shows the entropy per particle for different excitation amplitudes, ranging from no perturbation to
$A=$ 0.60$\upmu_0$, for several values of $t_{\rm hold}$.
The unperturbed cloud shows an approximately constant entropy per particle as a function of the time spent in the trap, as expected.
For a relatively small excitation, $A=$ 0.20$\upmu_0$, the entropy per particle is slightly higher than in the previous case, but with the same qualitative behavior.
Increasing the excitation amplitude even further, $A=$ 0.25 and 0.30$\upmu_0$, increases the entropy of the system, but still gives an approximately constant behavior as a function of the time.
For $A=$ 0.40$\upmu_0$, the system is at the threshold of becoming turbulent: the entropy per particle is higher than previous values, and the slope is more inclined for the initial times. For higher excitations, $A\geqslant$ 0.50$\upmu_0$ ,turbulence is fully developed. For $t_{\rm hold}\approx$ 45 ms, the entropy per particle experiences a sudden increase, which occurs after the particle cascade was established.

\begin{figure}[H]
\centering
\includegraphics[angle=-90,width=0.45\linewidth]{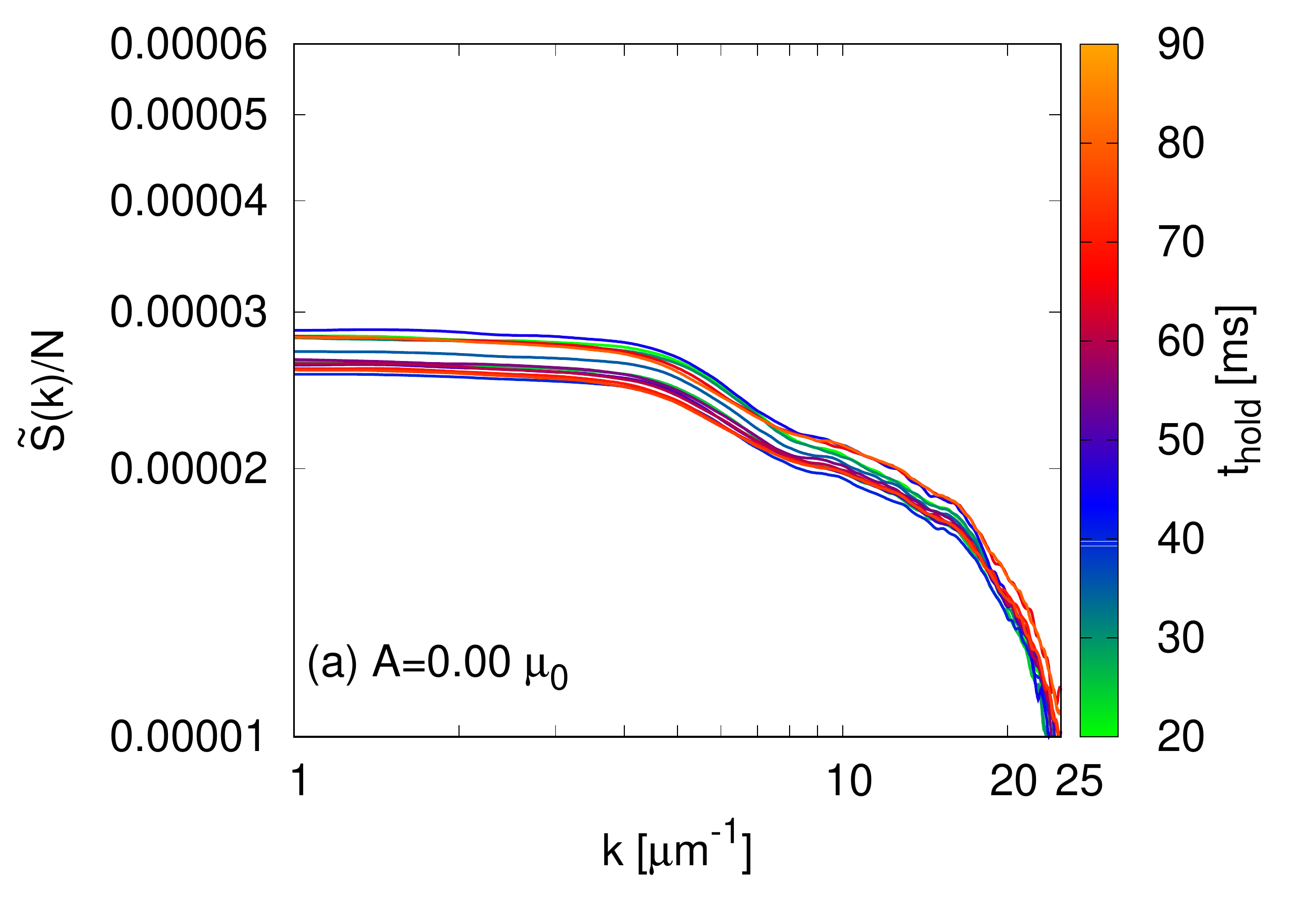}
\includegraphics[angle=-90,width=0.45\linewidth]{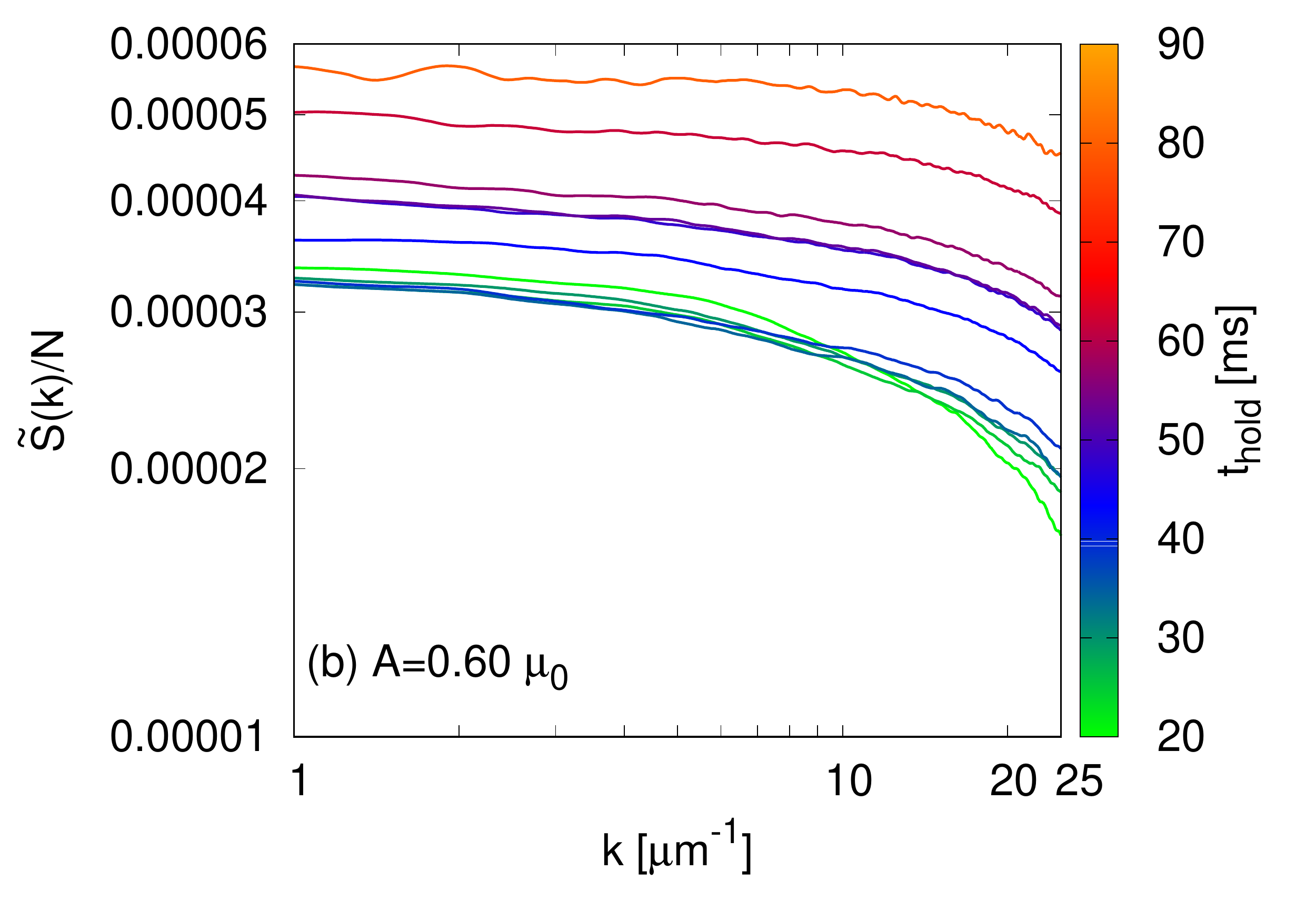}
\caption{Entropy per particle as a function of the momentum for an unperturbed BEC (\textbf{a}) and a turbulent cloud (\textbf{b}). Notice that the same scale was employed in both plots. The turbulent BEC experiences a sudden increase in the entropy per particle, for $t_{\rm hold}\geqslant$ 45 ms, and it reaches thermal equilibrium for long times, corresponding to a more flat distribution of entropy among the momentum~classes.
}
\label{fig:sofk}
\end{figure}

Much of our understanding of turbulent systems comes from functions computed in the momentum space, such is the example of the momentum distribution.
This is because they provide information on how particles and energy flow from one momentum class to the neighboring ones.
We~can define an entropy as a function of $k$ by recasting Equation~(\ref{eq:entropy}) as $S\equiv\sum_j \tilde{S}(k_j)$.
We show our results in Figure~\ref{fig:sofk} for $A=$ 0 and $A=$ 0.60$\upmu_0$, an unperturbed and a turbulent BEC, respectively.
The~ sudden increase in the entropy is also evident in Figure~\ref{fig:sofk}b, as the distributions for $t_{\rm hold}\leqslant$ 45 ms are close together and for longer times they continue to increase.
Thermalization occurs for long times for the turbulent BEC, and it corresponds to a more flat distribution.

The main mechanism of entropy generation, after the excitations were introduced in the sample, occurs by the interactions between them generating, in the case of vortices, Kelvin waves, and random distribution of fragments and rings, and, in the case of waves, a process of proliferation of waves that cascade up in momentum.
The generation of turbulence consists of converting the energy placed in an organized manner in the large length scales and migrating it to the small length scales, promoting~entropy formation.
In this sense, the capacity to produce excitations and promote their interaction is essential for the fast production of disorder and consequently, entropy.
Through this energy migration from large to small length scales, the cascade, a larger population of the sample is transferred to the high-momenta region. Over time, if a small number of excitations (low amplitudes) were introduced, there are few temporary occurrences to promote the disorder.
The dynamic with few excitations is slow, and even with a longer waiting time, it still does not produce enough interactions to promote the growth of the disorder. As a result, for low amplitudes, the growth of entropy is low over time. For excitation amplitudes up to values in the order of 0.30$\upmu_0$, the change in entropy over time is negligible.
When excitations are produced at the amplitude of 0.40$\upmu_0$, the rate of change in entropy begins to be more sensitive, demonstrating that sufficient excitations were generated for a high interaction rate and, therefore, a more significant generation of the disorder.

The final establishment of a cascade corresponds to an efficient transfer of energy from the largest length scales to the smallest scales and, thus, an efficient mechanism for the growth of entropy.
For~amplitudes of the order of 0.50$\upmu_0$ and above, this cascade establishment occurs not only due to the greater introduction of excitations in the sample, but also due to its rapid interaction. In general, the~establishment of turbulence is associated with this situation and, therefore, at the same time as the turbulence is established, entropy also has a higher growth rate.

%%%%%%%%%%%%%%%%%%%%%%%%%%%%%%%%%%%%%%%%%%
\section{Discussion and Conclusions}
\label{sec:conclusions}

In summary, we calculated the entropy of a BEC in regimes that go from
an unperturbed system up to a turbulent cloud, corresponding to a far-from-equilibrium state. We discussed how relatively small perturbations, the onset of turbulence, and the fully developed turbulent regime impact the momentum distribution and entropy of the system. In particular, we observed a sudden increase in the entropy following the establishment of a particle cascade.

These calculations were only possible because we derived an expression for the entropy, Equation~(\ref{eq:entropy}), with the data available from experiments in mind. The main assumption is that the expected value of the momentum distribution is well-known, i.e., the square of the amplitudes of the wave function are determined, whereas no information is available about the phase. Similar~considerations have been applied successfully in the field of quantum turbulence. For~example, the theory of wave turbulence \cite{Nazarenko2011} assumes a uniform random distribution for the phase of the different momentum classes.

It is also interesting to note that an expression for the entropy could also be derived from the kinetic equations that describe the non-equilibrium quasiparticle distribution function for a dilute inhomogeneous BEC \cite{Kirkpatrick1985a,Kirkpatrick1985b,Kirkpatrick1985c}.
If the same procedure was repeated with the assumptions, we employed in this work, namely the momentum distribution values are well-known and we have no knowledge of the phase, we would reach an expression for the entropy based on a microscopic description of the Bose gas.
Comparing the results that were obtained with both approaches could shed light on the role of microscopic interactions in developing the turbulent regime.

In contrast to experiments, the phase is accessible
in numerical simulations employing the Gross--Pitaevskii equation \cite{Tsubota2017}.
Performing simulations that reproduce the experimental conditions reported in this work would be illuminating in two main aspects.
First, the von Neumann entropy could be computed straightforwardly and the results as compared to our present work to test the validity of our assumptions. Second, it would be possible to distinguish between vortices and wave contributions to the turbulent regime, which would help to characterize the turbulence that we observe.

The identification of a power-law behavior of the momentum distribution has some intrinsic difficulties \cite{Madeira2020}, mainly because the region where it is observed is very narrow in BECs.
Hence,~other~approaches may provide useful insights into identifying and characterizing the turbulent regime.
We~believe that entropy and related quantities can be used to improve our understanding of quantum turbulence. It captures the essential features of the transition from equilibrium to a turbulent state and the changes that occur in between.

%%%%%%%%%%%%%%%%%%%%%%%%%%%%%%%%%%%%%%%%%%
\vspace{6pt} 

%%%%%%%%%%%%%%%%%%%%%%%%%%%%%%%%%%%%%%%%%%
\authorcontributions{
Conceptualization, F.E.A.d.S. and V.S.B.;
methodology, L.M. and F.E.A.d.S.;
software, L.M.;
validation, L.M. and A.D.G.-O.;
formal analysis, L.M.;
investigation, L.M. and A.D.G.-O.;
resources, V.S.B.;
data~curation, L.M. and A.D.G.-O.;
writing--original draft preparation, L.M.;
writing--review and editing, F.E.A.d.S.~and~V.S.B.;
visualization, L.M.;
supervision, V.S.B.;
project administration, L.M.;
funding acquisition, V.S.B.
All authors have read and agreed to the published version of the manuscript.}

%%%%%%%%%%%%%%%%%%%%%%%%%%%%%%%%%%%%%%%%%%
\funding{
This work was supported by
the S\~ao Paulo Research Foundation (FAPESP)
under the grants 2013/07276-1, 2014/50857-8, and 2018/09191-7, and by the
National Council for Scientific and Technological Development (CNPq)
under the grant 465360/2014-9.
FEAS thanks CNPq (Conselho Nacional de Desenvolvimento Cient\'ifico e
Tecnol\'ogico, National Council for Scientific
and Technological Development) for support through Bolsa de
produtividade em Pesquisa Grant No.
305586/2017-3.
}

%%%%%%%%%%%%%%%%%%%%%%%%%%%%%%%%%%%%%%%%%%
\acknowledgments{We thank
A.R. Fritsch, P.E.S. Tavares, and A. Cidrim
for the useful discussions.
}

%%%%%%%%%%%%%%%%%%%%%%%%%%%%%%%%%%%%%%%%%%
\conflictsofinterest{The authors declare no conflict of interest.} 

%%%%%%%%%%%%%%%%%%%%%%%%%%%%%%%%%%%%%%%%%%
%% optional
\abbreviations{The following abbreviations are used in this manuscript:\\

\noindent 
\begin{tabular}{@{}ll}
BEC & Bose-Einstein condensate\\
QUIC & Quadrupole-Ioffe configuration\\
TOF & Time-of-flight
\end{tabular}}

%%%%%%%%%%%%%%%%%%%%%%%%%%%%%%%%%%%%%%%%%%
\reftitle{References}
\bibliographystyle{mdpi}
\renewcommand\bibname{References}

%=====================================
% References, variant B: internal bibliography
%=====================================
%\begin{thebibliography}{999}
%% Reference 1
%\bibitem[Author1(year)]{ref-journal}
%Author1, T. The title of the cited article. {\em Journal Abbreviation} {\bf 2008}, {\em 10}, 142--149.
%% Reference 2
%\bibitem[Author2(year)]{ref-book}
%Author2, L. The title of the cited contribution. In {\em The Book Title}; Editor1, F., Editor2, A., Eds.; Publishing House: City, Country, 2007; pp. 32--58.
%\end{thebibliography}
%%%%%%%%%%%%%%%%%%%%%%%%%%%%%%%%%%%%%%%%%%
\end{document}